\documentclass[12pt]{article}
\usepackage{graphicx,epsfig}
\begin{document}

\hsize 14.0 cm

\centerline  {\Large {\bf Econophysics as conceived by  Meghnad Saha}}

\vskip 1.5 cm
           
\centerline {Bikas K. Chakrabarti}
     
\centerline {Saha Institute of Nuclear Physics, Kolkata 700064}

\centerline {S. N. Bose National Centre for Basic Science, Kolkata 7000106}

\centerline {Economic Research Unit, Indian Statistical Institute, Kolkata 700108}

\vskip 1.5 cm

\noindent {\bf Abstract:} We trace the initiative by Professor Meghnad
Saha to develop a (statistical) physics model of market economy and
his search for the mechanism to constrain the entropy maximized 
width of the income distribution in a society such that the spread
of inequality can be minimized.

\vskip 2 cm

\noindent {\bf I. Introduction:} Professor Meghnad Saha had been a great physicist  
and perhaps an even greater social scientist and reformer. It
is this second aspect, which is not  analysed often or 
highlighted  much in the literature about him. In this brief 
article, I would like to discuss his science-based attempts to 
model social dynamics and about his inspiring attitude 
towards social science in general.

\bigskip

\medskip

\noindent {\bf II. Saha's science-based approaches for social problems:}
There are several writings about him which talks about Professor
Saha's deep interest and involvement in matters of our day-to-day
social concern. Let me quote from a book and a recent article on
Saha, as examples: Dilip M. Salwi, in his book titled `Meghnad Saha: 
Scientist with a Social Mission' [1] writes ``{\small {\it... Meanwhile 
in 1943, Damodar river went into spate, its
flood waters even surrounded Kolkata and cut it off from 
rest of India. ... Saha took upon himself the task of
studying the flood problem of Damodar river in totality 
because it also caused soil erosion and siltation.  In 
addition to the study of the topography of the region,
the annual rainfall at various spots, etc., he set up a
small hydraulic laboratory in the college to simulate the
actual conditions and understand the problem. He also
conducted a first-hand survey of the region and even studied
the various flood control measures of river systems of the
western world. On the basis of his painstaking research, he
wrote a series of articles in his own newly launched monthly 
Science and Culture. ... To sensitize his colleagues, friends,
and students to the menace of floods, he also set up a model
of Damodar river system in the corridor of Science College!}}".
Pramod V. Naik in his article [2] writes  ``{\small {\it... Meghnad 
Saha  (1893-1956) ...  was of the firm 
opinion that  in a country like India, 
the problems of food, clothing, eradication of 
poverty, education and technological progress can 
be solved only with proper planning, using science 
and technology. ... While thinking about various 
issues of national importance, Saha realized the
need for a scientific periodical, ... and 
founded a monthly titled Science
and Culture. ... The range of
topics in editorials and general articles
was amazing; for example, the need for a
hydraulic research laboratory, irrigation
research in India, planning for the
Damodar Valley, ..., public supply of
electricity in India, national fuel policy,
... mineral sources and
mineral policy, problem of industrial
development in India, automobile industry
in India, industrial research and Indian
industry, industrial policy of the Planning 
Commission, ... and so on. Every article shows
his inner urge for the goal of national
reconstruction. ... show his direct or 
indirect role as a social thinker and 
reformer.}}". See also the paper by Vasant 
Natarajan is this special issue.

\bigskip

\medskip

\noindent {\bf III. Saha and Srivastava's Kinetic Theory for ideal
gases with `atoms' and `social atoms':} No wonder, Professor Saha 
had  thought deeply about 
the scientific foundation of many social issues. In
particular the omnipresence of social (income or 
wealth) inequalities in any society throughout the  
history of mankind must have caught his serious 
attention. Like all the  philosophers through ages,
 and also as a born socialist, he must have been 
pained to see social inequalities cutting across 
the history and the globe at any point of time. Yet, his confidence
in science, statistical physics in particular, perhaps
convinced him that `entropy maximization' principle
must be at work in the `many-body' system 
like the society or a market, and some amount of inequality might be
inevitable! One Robinson Crusoe in an island can not
develop a market or a society. A  typical thermodynamic
system, like a gas, contains Avogadro number ($N$ about
$10^{23}$) of atoms (or molecules). Compared to this, 
the number ($N$) of `social atoms' or agents in any market
or society is of course  very small (say, about $10^2$ for a
village market to about $10^{9}$ in a global market). 
Still such many-body dynamical systems  are statistical
in nature and statistical physical principles should be applicable.
 In the famous text book `Treatise on Heat'
[3],  written together with Professor Biswambhar Nath Srivastava, 
the students were encouraged, in the section on
Maxwell-Boltzmann distribution in kinetic theory of ideal
gas,  to think about applying kinetic theory to the 
market and find the income distribution in a society,
which maximizes the entropy (for stochastic market transactions). 
It may be noted, this point was not put at the
end of the chapter. It was put right within the text
of the section on energy distribution in ideal gas. It provokes the student 
to search for the equivalent money or income (wealth) distribution in
a stochastic market when money is conserved in each 
of the transactions.   

\begin{figure*}[htb]
\begin{center}
\fbox{\includegraphics[width=10.0cm, angle=-89]{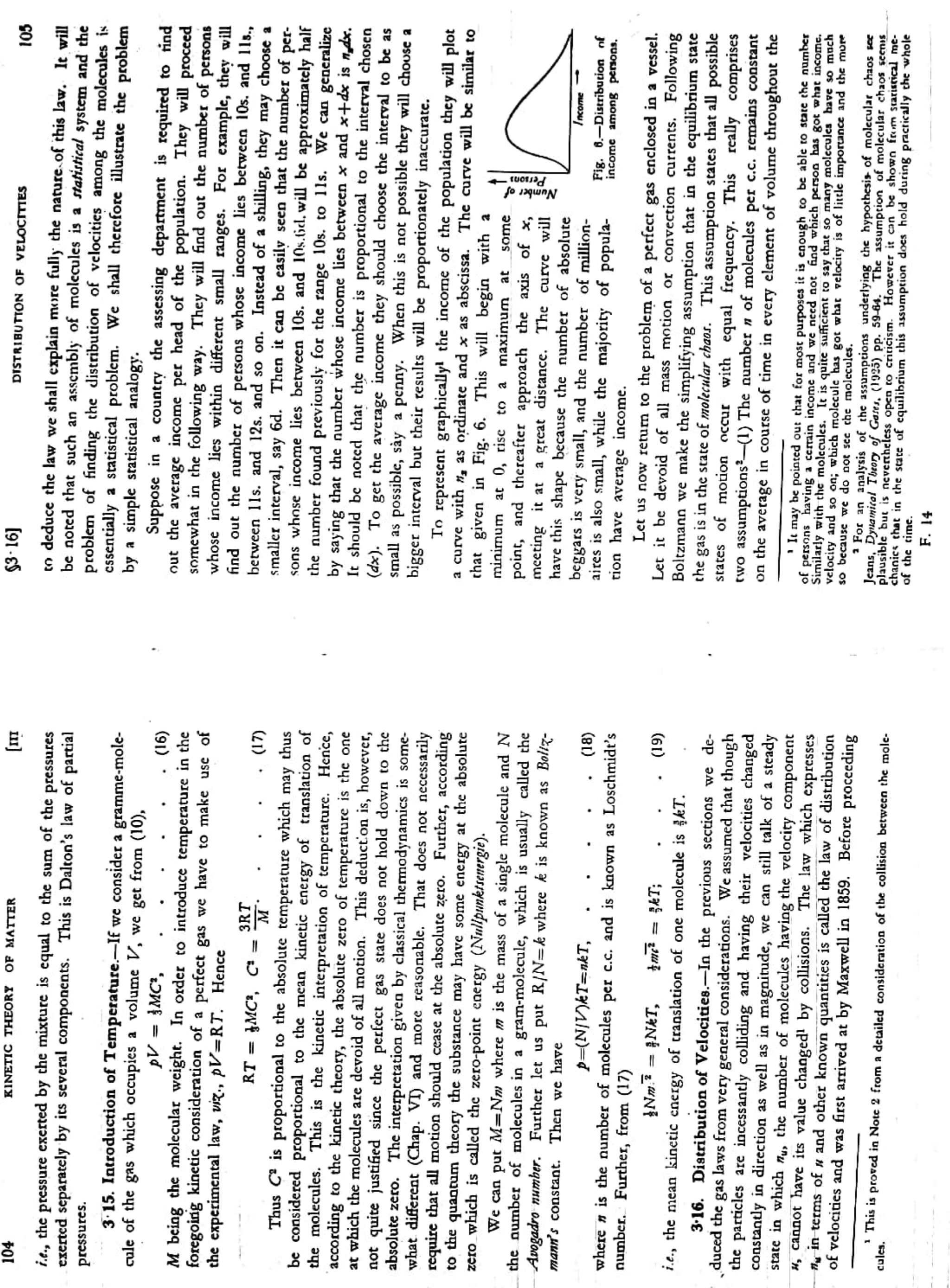}}
%\fbox{\includegraphics[width=13.0cm]{Saha_Srivastava2.pdf}}
\end{center}
\caption{ Photocopy of the pages 104 and 105 of Ref. [3], where
the authors urge the students to use kinetic theory to get
the Gamma-like income distribution indicated in Fig. 6 of
page 105.}
\label{saha_book}
\end{figure*}

One can   present the
derivation of the energy distribution in an ideal gas in
equilibrium at a temperature $T$ as follows: If 
$n(\epsilon)$ represents the number density of particles
(atoms or molecules of a gas) having energy $\epsilon$,
then $ n(\epsilon) d\epsilon =  g(\epsilon) f(\epsilon , T)
d\epsilon $. Here $g(\epsilon) d\epsilon $ denotes the 
denotes the `density of states' giving the number of
dynamical states possible for free particles of the
gas, having kinetic energy between $ \epsilon$ and $\epsilon 
+ d\epsilon$ (as dictated by the different momentum vectors
${\vec{p}}$ corresponding to the same kinetic
energy: $\epsilon  = {|\vec{p}|}^2 /2w$,
where $w$ denotes the mass of the particles). Since the 
momentum ${\vec {p}} $ is a three dimensional vector,
$g(\epsilon)d\epsilon \sim |\vec {p}|^2 d|\vec {p}|
\sim |\vec {p}||\vec {p}|d|\vec {p}| \sim \sqrt \epsilon d\epsilon$.
This is obtained purely from mechanics. For completely 
stochastic (ergodic) many-body dynamics or energy exchanges, 
maintaining the the energy conservation, the energy distribution 
function $f(\epsilon) (\equiv f(\epsilon, T))$ 
should satisfy $f(\epsilon_1) f(\epsilon_2)
= f(\epsilon_1 + \epsilon_2)$ for any arbitrary $\epsilon_1$
and $\epsilon_2$. This suggests $f(\epsilon) \sim exp (-\epsilon/ \Delta)$,
where the factor $\Delta$ can be identified from the equation
of state for gas, as discussed later (=  {\large$\kappa$}$T$,
{\large $\kappa$} denoting the Boltzmann constant). This gives
 $n(\epsilon) = g(\epsilon)f(\epsilon) 
 \sim {\sqrt \epsilon} exp(-\epsilon/\Delta)$. Knowing 
this $n(\epsilon)$, one can estimate the average pressure $P$
the gas exerts on the walls of the container having volume $V$
at equilibrium temperature $T$ and compare with the ideal gas
equation of state $PV = N${\large$\kappa$}$T$. 
 The pressure can 
be estimated from the average rate of momentum transfer by the atoms on 
the container wall and one can compare with that
obtained from the above-mentioned equation of state and 
identify different quantities; in particular, one idenifies 
$\Delta =$ {\large$\kappa$}$T$.

Saha and Srivastava's chosen example in this section (see Fig.1)
indicated to the students that since in the market money is 
conserved as no one can print
money or destroy money (will end-up in jail in both
cases) and the exchange of money in the market is 
completely random, one would again expect, for any
buyer-seller transaction in the market,  $f(m_1)f(m_2) =
f(m_1 + m_2)$, where $f(m)$ denotes the equilibrium or steady state 
distribution of money
$m$ among the agents in the market. This, in a similar way, suggests
$f(m) \sim exp(-m/\sigma)$, where $\sigma$ is a constant. Since
there can not be any equivalent  of the particle momentum vector for the
agents in the market, the density of states $g(m)$ here is a constant
(each value of $m$ corresponds to one market state). Hence, the
number density $n(m)$ of agents having money $m$ will be
given by $n(m) = C exp(-m/\sigma)$, where $C$ is another constant. 
One must also have $\int _0^{\infty} n(m) dm = N$, the total
number of agents in the market and $\int_o^{\infty} mn(m)dm = M$, the 
total money in circulation in the country. This gives, $C = 1/\sigma $
and $\sigma = M/N$, the average available money per agent in this
closed-economy traders' market (as no growth, migration of labourers, etc. 
are considered). This gives exponentially decaying (or Gibbs-like) distribution
of money  in the market (unlike the Maxwell-Boltzman or
Gamma distribution of energy in the ideal gas), where most of the 
people become pauper ($n(m)$ is maximum at $m = 0$). 
They asked the students to investigate, what could
make the distribution more like Gamma distribution, as seemed to
them to be observed phenomenon in most of the societies. Any understanding
towards  that would help to identify and develop public policies to
reduce the extent  of inequalities in the society. It may be mentioned 
here that although the statistical aspects of societies, markets in 
particular, had been addressed in some speculative papers earlier,
the 1931 book `Treatise of Heat' [3] is the first ever text book addressing
the question in the context of kinetic theory of gas and pleads for
the search for a solution from statistical physics!     

\bigskip

\medskip

\noindent {\bf IV. Saha's Econophyics contribution as noted in some
on-line enclyopedias:} In the entry on `History of Indian School of 
Econophysics', in the online encyclopedia (Hmolpedia) [4]
Libb Thims (2016) writes: ``{\small {\it ... In 1931, 
Indian astrophysicist Meghnad 
Saha (1893-1956), an atheist, in his Treatise on Heat,
co-authored with B.N. Srivastava, explained the Maxwell-
Boltzmann distribution of molecular 
velocities according to kinetic theory in terms of the 
wealth distributions in society [Saha, Meghnad and 
Srivastava, B. N. (1931), Treatise on Heat (pg. 105), 
The Indian Press Ltd. (1931)]. Indian physicist Prasanta 
Mahalanobis (1893-1972), interested in Karl Pearson 
stylized biometrics, founded in 1931 the Indian Statistical 
Institute for developing physical and statistical 
models for social dynamics. In 1938, Saha began to 
form the Saha Institute of Nuclear Physics in Kolkata, 
India.}}".  The Wikipedia entry  (June, 2018) on  `Kinetic 
exchange models of markets' [5]
reads: ``{\small {\it ... Kinetic
exchange models are multi-agent dynamic 
models inspired by the statistical physics of 
energy distribution, which try to explain the 
robust and universal features of income/wealth 
distributions. Understanding the distributions 
of income and  wealth in an economy has been a 
classic problem in economics for more than a 
hundred years. Today it is one of the main 
branches of econophysics. ... In 1897, Vilfredo 
Pareto first found a universal feature in the 
distribution of wealth. After that, with some 
notable exceptions, this field had been dormant 
for many decades, although accurate data 
had been accumulated over this period. Considerable 
investigations with the real data during the last 
fifteen years revealed that the tail (typically 5 
to 10 percent of agents in any country) of the 
income/wealth distribution indeed follows a 
power law. However, the majority of the population 
(i.e., the low-income population) follows a 
different distribution which is debated to be 
either Gibbs or log-normal. ...  Since the 
distributions of income/wealth are the 
results of the interaction among many heterogeneous 
agents, there is an analogy with statistical mechanics, 
where many particles interact. This similarity was 
noted by Meghnad Saha and B. N. Srivastava in 1931
[Saha, M. \& Srivastava, B. N., A Treatise on Heat. 
Indian Press (Allahabad; 1931). p. 105], and thirty 
years later by Benoit Mandelbrot\footnote{Mandelbrot, B. B.,
The Pareto-Levy law and the distribution of income. 
International Economic Review. {\bf 1} (1960) pp. 79-106;
Indeed he wrote: “There is a great temptation
to consider the exchanges of money which occur in 
economic interaction as analogous to the exchanges 
of energy which occur in physical shocks between 
molecules. In the loosest possible terms, both kinds 
of interactions should lead to similar states of 
equilibrium. That is, one should be able to explain 
the law of income distribution by a model similar 
to that used in statistical thermodynamics:
many authors have done so explicitly, and all the 
others of whom we know have done so implicitly.”}. 
In 1986, an elementary version of the stochastic 
exchange model was first proposed by Jhon Angle 
[Angle, J. (1986). The surplus theory of social 
stratification  and the size distribution of personal 
wealth,  Social Forces, {\bf 65} (1986), pp. 293-326].}}"

\bigskip

\medskip

\noindent {\bf V. Brief discussions on recent developments:}
Saha and Srivastava had been in search of the mechanism
that will ensure the initial fall (as in the Maxwell-Boltzmann
distribution $n(\epsilon)$ of energy for the Newtonian particles 
in the gas, due to the particle momentum count of the density of 
states) in the Gibbs-like money distribution $n(m)$ in the 
market\footnote{For a recent account on the extent and
consequences of this Gibbs-like distribution in monetary econophysics,
see: V. M. Yakovenko and J. Barkley Rosser, Statistical mechanics of
money, Reviews of Modern Physics, {\bf 81}, 1703-1725 (2009)}.
They inspired the students to investigate the plausible reasons for this
`Gamma' distribution like initial fall (from the scenario where
most people are pauper or $n(m)$ is maximum at $m = 0$), which they rightly
conjectured to be the realistic form of the money distribution
in any society.  There was, however, another aspect of the distribution
$n(m)$ in any society. This was first observed and formulated in 1896 by 
Vilfredo Pareto (Engineer cum economist/sociologist of the
Polytechnic University of Turin),  known as Pareto law [6]. This law  
states that for the number of truly rich people in any society $n(m)$
does not fall off exponentially (as in the kinetic theory model 
indicated above), rather fall  off with a `fat tail'
having  an `universal' power law decay: 
$n(m) \sim m^{-\alpha}$, with $\alpha$-value in the range 2 - 3.

The market model considered above by Saha and Srivastava could
at best be described as a trading market (with fixed value of $N$ and $M$)
as it does not accommodate in this limit any  growth, industrial
development or migration of labours etc. If one consults a standard
economics text book and looks for the discussions on trading market,
one can not miss the discussions on `saving propensity'! This quantity
characterizes the agents or traders by the fraction of their money
holding at that instant they save before going  in for a trade. Though, there
can be some fluctuations in the value of this fraction from trader to trader, 
and can even change with time  for any individual
trader, their respective 
time-averaged values can characterize different traders' 
attitudes. Even the traders from different countries seem to have 
clear differences in their average saving propensities!
In kinetic theory, the atoms do not have any identity, while the
social atoms seem to maintain their identity through their
respective propensity to save in any transaction or trade
(and also maintain the memory of the past transactions). This
means, the kinetic exchanges in such trading markets are
money conserving but non-Markovian exchanges.

In view of the above-mentioned observation by the economists
for a trading market, and the realisation that kinetic theory of
gas does not allow for any `saving' of energy of any atom before
its exchange with  another, it was proposed (see e.g., [7] and 
the references therein) that
the kinetic theory of market may be extended a bit by allowing 
saving propensity of the `social atoms', and the money conserving 
exchange  kinetic equations can be rewritten. It was obvious that, say
with the same saving propensity of all the agents in the market, the
Gibbs distribution will collapse to a Gamma-like distribution: Once
any one having a finite saving propensity  acquires some money, to 
become a pauper he/she will have to lose successively 
in the market in every one of the  later trades or 
transactions. The probability of such an event, for any finite 
saving propensity value of the agent, is infinitesimal. This is
a nonperturbative result: $n(0)$ will drop down to zero from its
maximum for any non-zero saving propensity of the agents, in the
steady state. The
most probable income per agent in the market will shift from zero to
a value dependent on the saving propensity of the agents. It also 
suggests immediately that in a market with people having different or
random values of saving propensity, there would be a net flow of 
money towards people with higher saving propensities and in the steady sate, 
a robust power law tail for the distribution, $n(m) \sim m^{-2} ~$ 
($m \rightarrow \infty$), will ensue for most of the non-singular 
probabilities of saving propensity among the population [7].
Many important developments and modifications on this kinetic
exchange model, pioneered by Saha and Srivastava in [3] (see also [7]),
have taken place since then, both in the mathematical physics
(see e.g., [8]) and in economics (see e.g., [9]).

\vskip 1 cm

\noindent {\bf Concluding Remarks:} Apart from being one of the pioneering 
astrophysicists of our time, Professor Meghnad Saha had been a passionate
and enthusiastic thinker in social science. His deep conviction
about scientific approaches to social phenomena  convinced him about
the applicability of  the laws physics, of statistical physics in 
particular, to social sciences. In view of the inherent many-body stochastic 
dynamical features of the markets,  he realised that the
entropy maximizing principle will
not support a steady state  narrow  income or money 
distribution among the traders. As a diehard socialist, this 
conclusion must have been quite painful to
digest. He was thus in search of the tuning parameters
for controlling the width of the money distribution $n(\epsilon)$, around
the most probable value of income (ideally the average money
per agent $M/N$ in
circulation in the market). Saha-Srivastava's kinetic exchange
model for a traders' market suggested that most traders will beecome
pauper in the steady state
($n(m)$ will be maximum at
$m = 0$). This comes due to the limitation of their model (where
the social atoms can lose its entire accumulated money in the
next trade with another, as in the case of real atoms in the gas) and also
such money distributions  do not quite compare either with the 
observational results in societies. They had,
however, every confidence on the students, whom they expected
to think about it and find eventually a  proper solution. As
mentioned above in section V, one such solution 
had been achieved later by adding finite saving 
propensities of the social atoms or traders in the Saha-Srivastava model, 
following the observations of economists in
the trading markets.  This additional feature  in the model 
naturally gives Gamma-like income distribution with tunable
dispersion of inequality near the most probable income and also 
induces the eventual Pareto-like slowly decaying
numbers (fat tail distribution) of the  ultra-rich people or 
traders in the society.

\vskip 1 cm

\noindent {\bf Acknowledgement:} I am grateful to Abhik Basu, 
Arnab Chatterjee,  Kishore Dash, Amrita Ghosh,
 Libb Thims and Sudip Mukherjee
for important comments, criticims and communications.   

\vskip 2 cm

\leftline {\bf References:}

\bigskip

\noindent [1] D. M. Salwi,  {\it Meghnad Saha: Scientist with a 
Social Mission}, Rupa \& Co., New Delhi (2002), pp. 49-51

\medskip

\noindent [2] P. V. Naik, `Meghnad Saha and his contributions',
Current Science, Vol. 111(1),  10 July 
(2016), pp. 217-218

\medskip

\noindent [3]  M. N. Saha and B. N. Srivastava, {\it A Treatise on Heat.}
Indian Press (Allahabad; 1931), p. 105 

\medskip

\noindent [4] L. Thims, `History of Indian School of Econophysics', in
free online Encyclopedia of Human Thermodynamics (Hmolpedia; as of June 2018):

\noindent \verb|http://www.eoht.info/page/Indian+school+of+econophysics|

\medskip 

\noindent [5] Wikipedia entry  (as of June, 2018) on  `Kinetic
exchange models of markets', in free online Encyclopedia:

\noindent \verb|https://en.wikipedia.org/wiki/Kinetic_exchange_models_of_markets|

\medskip

\noindent [6] V. Pareto, {\it Cours d’Economie Politique}, Droz, Geneva (1896)

\medskip

\noindent [7] B. K. Chakrabarti, A. Chakraborti, S. R. Chakravarty and
A. Chatterjee, {\it Econophysics of Income \& Wealth Distributions}, 
Cambridge University Press, Cambridge (2013)

\medskip 

\noindent [8] L. Pareschi and G. Toscani, {\it Interacting Multiagent 
Systems}, Oxford University Press, Oxford (2014)

\medskip

\noindent [9]  M. Shubik and E. Smith, {\it The Guidance of an 
Enterprise Economy}, MIT Press, Cambridge, Massachusetts (2016)

\end{document}